# A study on the thermal conductance of interface between dissimilar metals

Dian Li[1], Joseph Feser[1]

[1] *Department of Mechanical Engineering, University of Delaware, 19716, DE, USA*

**ABSTRACT**

Whether diffuse mismatch model for electrons (DMMe) hold true in more general cases remains largely unexplored, especially in cases where at least one material does not behave like a free- electron metal and/or the interface is smooth enough to allow non-diffuse transmission of electrons. In this study, DMMe was proposed to predict the thermal conductance of metal-metal interfaces. A set of aluminum-X samples (X = Cu, Ag, Fe, Ni) were grown and the time domain thermoreflectance (TDTR) technique was used to measure the metal-metal interface conductance. It was then compared to the two variantsof the DMMe - using both a crude theory based on free-electron metals and using accurate band structures provided by density functional theory.

## 1. Introduction

Heat transfer in solids at nanoscale is governed by the heat transport of its energy carriers, i.e. electrons and phonons [1, 2, 3, 4, 5, 6]. In the interface between two solid layers, heat transfer happens via the collisions between its energy carries, with the gradient of different energy distribution because of temperature difference [7, 8]. In bulk metals, energy is transported primarily by electron diffusion [6, 9, 10, 11, 12, 13, 14]. There are only two works, which investigated the thermal transport efficiency of metal-metal interfacial layers. Recently, Ruan *et al.* [15] have used density functional theory to estimate the phonon contribution to thermal conductivity for aluminum is just 5.8 W/m-K, which is consistent estimation of the electronic contribution of thermal conductivity using the Weidemann-Franz law (using $\rho = 2.8 \times 10^{-8}\Omega$-m gives $k_e = L_0 T/\rho = 260$ W/m-K) [10, 11, 16, 17, 18]. At metal-metal interfaces, where electrons can transmit from one side to the other, electrons are also thought to be the primary carriers of heat. Moreover, Gundrum *et al.* [19] measured the interfacial conductance between an individual Al-Cu interface using TDTR. At room temperature, they reported an extremely large interfacial thermal conductance value of 4,000 MW/m²; for reference, the highest ther-mal interface conductance for a metal-dielectric interface ever measured is currently 700 MW/m²-K for an epitaxial TiN-MgO interface, and most non-epitaxial metal- dielectric interfaces lie between 20 - 200 MW/m²-K. They also proposed a theory that they claim is consistent with their data based a modification of the diffuse mismatch model, originally developed to model phonon transport across disordered/rough interfaces [20]. The primary assumption of the diffuse mismatch model is that when a carrier reaches an interface, it scatters and loses memory of which side of the interface it originated from; the mathematics of this theory will be discussed in the next section. In addition, Gundrum observed a continual and nearly linear increase in the interface conductance with temperature over the range 77K-300K. For reference, metal-dielectric interface conductance generally follows the same temperature dependence as phonon heat capacity, transitioning from $T^3$ toward $T^0$ dependence (i.e. flat) near room temperature in most metals. Though Gundrum's work shows excellent agreement between theory and experiment, there remain several open questions. Al-Cu can form a lot of stable compounds at room temperature ($Al_4Cu_9$ in particular) that could enhance the chemical roughness of the interface, so it is not clear that if the diffuse mismatch model holds for the Al-Cu, that this should be true for other systems. In addition, Gundrum made some questionable assumptions during his implementation of the diffuse mismatch model, most notably (1) treating both metals as having parabolic bands/a spherical Fermi surface and (2) using experimental values of the heat capacities rather than the electronic density of states directly. It is unclear whether either of these two assumptions should be correct for Al or Cu, and they would certainly not be applicable for most transition metals.

In this study, we prepare a variety of samples with metal-metal interfaces, measure their thermal interface conductance using time-domain thermoreflectance, and compare the experimental results to various theoretical estimates. Since aluminum is known to be a good thermoreflectance transducer, we have designed the experiments so that aluminum always serves as the topmost layer of our samples and thus is the only surface that interacts with our laser.

## 2. Methods

### 2.1. Sensitivity simulation and analysis

The details of our sample designs are motivated by a TDTR sensitivity analysis that we have made based on the known bulk properties of the constituent metals and expected interfacial conductance based on Ref. [21]. In Gundrum's original work, they measured interface conductance of an Al/Cu interface using two samples, that is, 50nm Al (top)/315nm Cu(middle)/ Sapphire (substrate) and 50nm Al (top)/1570nm Cu(middle)/ Sapphire (substrate) thin. The reason for having two samples is that the thermal penetration depth of Cu at a modulation frequency of $f$ =10 MHz (where the measurement was done) is about $L_p \approx$ 1300 nm, where the penetration depth is calculates as

$$L_P = \sqrt{\frac{D}{\pi f}} \qquad (1)$$

where $D$ is thermal diffusivity of the metal and $f$ is the modulation frequency of the incident laser. Thus, in the thin Cu sample, the temperature profile within the sample protrudes very far into the Sapphire, and includes the effects of both the Al/Cu interface resistance and the thermal interface conductance of the Cu/Sapphire interface. The Cu/Sapphire interface resistance is about one order of magnitude larger than the Al/Cu interface resistance, making the Cu/Al unmeasurable in this case. On the other hand, the thick Cu sample has penetration depth comparable to the thickness so that the temperature gradients in the sample are mostly (but not entirely) confined above the Cu/Sapphire interface. Using the Cu/Al interface conductance (measured in the first sample), the Al/Cu interface conductance can be extracted.

Based on the two-sample strategy employed by Gundrum, we have made a list of the estimated thermal penetration depths (and thus minimum sample thicknesses necessary) for various metal underlayers in Table 1. However, having a sample thick-ness like the thermal penetration depth is not a sufficient condition for being able to measure the thermal interface conductance. Metal-metal interfaces have very large conductance (4-12 GW/m²-K) such that their resistance (1/$G$) can be small com- pared to the thermal resistance associated with bulk conduction $R_{cond} \approx L_p/k$ inside the metal. Thus, for metals with low thermal conductivity and on the higher end of interface conductance, the bulk conduction obscures information about the interface conductance. Based on this logic, we categorized the metals in Table 1 according to the relative ratio between the interface resistance and the bulk conduction resistance.

2.2 Metal-metal superlattice structure interfacial layers

In order to improve the sensitivity of metal-metal interface measurements be- yond what is possible using the two-sample approach, and to lower the deposition thickness (saving time and costly material), we have come up with an innovative way to deposit the that ensures the interfacial thermal conductance dominates the overall thermal conductance of the metal layers. The idea is to create superlattices using the two metals; the finer the superlattice spacing, the great the interface density and the more sensitive we become to interface properties. In this configuration, if we study the unit cell of the superlattice (Figure 1), we see that the overall thermal conductance of a unit cell can be calculated as

$$\frac{1}{G_{total}} = \frac{L_A + L_B}{\kappa_{eff}} = \frac{L_A}{\kappa_A} + \frac{L_B}{\kappa_B} + \frac{2}{G_{AB}} \qquad (2)$$

where the superlattice unit cell consists of one layer of metal A and one layer of metal B with two interfaces. In Figure 1, a unit cell of the superlattice system is in $L_A + L_B$ length. Rearranging this, the effective thermal conductivity that would be seen by an experiment where the temperature gradient spans many layers would be

$$\kappa_{eff} = \frac{G\kappa_A\kappa_B(L_A + L_B)}{G(\kappa_A L_B + \kappa_B L_A) + 2\kappa_A\kappa_B} \qquad (3)$$

In all cases, we have chosen equal thickness for both metal layers, i.e. $L_A = L_B$. This has no influence on the accuracy of the analysis and makes the calculation easier. Due to the stacked interface resistances, the penetration depth in superlattice configurations are much thinner than for a bilayer deposition. The effective thermal diffusivity of the the superlattice sample can be calculated as

$$D_{eff} = \frac{4\kappa_{eff}}{(C_{pA} + C_{pB})(d_A + d_B)} \quad (4)$$

where $C_p$ and $d$ refer to specific heat and density respectively. We can then use equation Eq. (1) to calculate the thermal penetration depth of the superlattice. For example, for a superlattice combination of Al ($\kappa$ = 180Wm$^{-1}$K$^{-1}$) and Cu ($\kappa$ = 370Wm$^{-1}$K$^{-1}$) with a 5nm thickness for each layer, its effective thermal conductivity is 18.47 Wm$^{-1}$K$^{-1}$. The penetration depth of this superlattice sample at 12.6 MHz periodic laser pulse heating is 350 nm. This value is much smaller than the 1675 nm for normal deposition of Cu to measure interfacial conductance. Since each Al/Cu bilayer is 10 nm thick, we need repeat Al/Cu deposition 35 times to keep temperature gradients confined to the metal superlattice. Finally, we still use a 50 nm Al top transducer layer. This sample design greatly improves the sensitivity of our experiments to interfacial conductance and decreased workload of deposition. For the repetition of the superlattice deposition, instead of manually switching power and activated target for metal coating, this process can be automated in our sputtering chamber using a programming interfaceand a series of timed pneumatic shutters for each target. In the table below (Table 3), thermal penetrations of the samples to be made are listed. Once the effective thermal conductivity of the superlattice configuration is determined, we can back-calculate the $G$ values using Eq. (2), which results in

$$G_{AB} = 2(\frac{L_A + L_B}{\kappa_{eff}} - \frac{L_A}{\kappa_A} - \frac{L_B}{\kappa_B})^{-1} \quad (5)$$

2.3. Metal candidate choices and theoretical expectations

For all of the different metals we have used to form interfaces with aluminum, we have performed basic estimates of the expected thermal interface conductance, based on their electronic band structure. In Gundrum's original paper [21], it is shown that the metal interfacial thermal conductance depends on the two metals' product of electronic heat capacity per unit volume ($\gamma$) and their Fermi velocities, $v_F$.

$$Z = \gamma T v_f \quad (6)$$

For free electrons one finds that,

$$\gamma T = \frac{\pi^2}{2} n k_B \frac{T}{T_f} \quad (7)$$

where $n$ stands for the free electron density of a metal, $\kappa_B$ is the Boltzmann constant, and $T_f$ represents the Fermi temperature of the metal (normally in a magnitude of $10^5 K$). The electronic heat capacity of a metal always has a linear relationship with its temperature when $T$ is much less than $T_F$. The volumetric free electron density $n$, can be estimated as

$$n = \frac{N_A \rho}{M_A} \quad (8)$$

where $N_A$ is Avogadro's constant, $6.02 \times 10^{23}$ mol$^{-1}$, $\rho$ the density of the metal (kg/m$^3$) and $M_A$ the atomic mass. Using this theory, we calculated the expected interfacial conductance for five metals, Cu, Ag, Sn, Fe and Ni, listed in Table 2 below, which we also seek to perform measurements on. As noted earlier, for some types of metals, the heat transfer process is not carried out by 'free electrons,' and there is a strong electron-phonon coupling and other non-isotropic band structure considerations that might make the free electron treatment inaccurate. With the calculated interfacial conductance, we deposit superlattice layers on sapphire substrates, along with a top Al transducer, to measure the experimental interfacial conductance of Al/X layer (X stands for the target metal). The sapphire substrate is cut from 900 $\mu$m thick sapphire wafer. We also use 500μm silicon substrates for the same depositions so that the samples can be measured with scanning electron microscope (SEM) due to the electrically conductive ability of silicon. To simplify Eq (1), we may also know the penetration depth of the superlattice metal layer piles.

$$L_P = 2\sqrt{\frac{4\kappa_{eff}}{(C_{pA}+C_{pB})(d_A+d_B)\pi f}} \tag{9}$$

For each Al/X material, we attempted to make two samples where the thickness of each layer in a superlattice unit cell was 5 nm or 10 nm. The thinner each layer is, the more dominant interfacial conductance becomes in terms of effective thermal properties. However, we limit the minimum thickness to 5 nm to leave enough space to allow some interdiffusion each interface. If the thickness is too small, the metal interfaces will interdiffuse each other and there might not be a well-defined interface. For each sample, we deposit a superlattice structure slightly thicker than the expected penetration depth to ensure the accuracy of the measurement. Note that since the interface conductance is not known *a priori*, the expected penetration depth is based on an educated guess from free electron theory; if the interface conductance turns out to be much greater than expected, then the real penetration depth will be larger, and the measurements become less accurate. Below is the table of penetration depth of superlattice configurations as well as the choice of superlattice bilayer number.

2.4. Metal-metal superlattice deposition

In order to produce the superlattice samples for TDTR measurement, we first determined the deposition rate of all the metals and their physical properties (note that the properties of a thin film are often not the same as the properties of bulk materials). We placed an Al target as a permanent source as it is one of the metal components of every superlattice structure sample. To provide a neat and uncontaminated substrate surface, we use piranha etch solution (a liquid mixture with 3:1 of concentrated sulfuric acid and 30% hydrogen peroxide solution) to clean the sapphire substrates. This effectively dissolves or removes organic particles. We use distilled or deionized water to wash the substrate and finally dry the substrates with Nitrogen gas flow.

We used radio frequency (RF) magnetron sputtering for the aluminum deposition. The deposition rate of aluminum in the condition of 20 sccm and 6 mTorr with a 195W RF output power is approximately 5 nm/min. For each 5 nm Al superlattice layer, Al target is powered 1 min per period. It is the first metal ahead of the other metal to be powered to adhere onto the clean substrate. Al is commonly in good adhesion to sapphire substrate. This, in one way, allows for the firm adhesion of the superlattice structures on the sapphire substrates. In another way, depositing aluminum first makes the deposition of superlattice end with the other metal, which is convenient for acting as a TDTR transducer. For the superlattice deposition, we have used a programmed procedure to stack the superlattice multilayers. The program repeats the deposition with a set period, which consists of an Al layer deposition and a layer of the other metal. We preset the parameters of the deposition, such as Ar flow, pressure, power, and shutter opening time (deposition time) to formulate all the conditions. Then we set the cycle number, which determines how many bilayers the targeted superlattice structure will have. The program automatically deposits superlattice films with the desired requirements. We place all non-magnetic metal targets in direct current slot (DC) for deposition. For copper, at the same physical condition with 39W DC power, it is sputtered at 5 nm/min. Copper has great difficulty in adhering to sapphire as it cannot formvery strong atomic bonds with sapphire. In Gumdrum's experiments, they sputtered and titanium layer onto the sapphire surface to boost the adhesiveness of the deposited film. In our experiment, if the substrates are thoroughly cleaned by piranha solution and is firstly coated with Al, there were no issues with adhesion of superlattice films.

2.5. Thermal conductivity measurement of metal targets

It is crucial to know the relatively accurate values of thermal conductivity of the metal targets. Here, we use $SiO_2$ substrates to deposit metal on it. BK7 glass is also a good option. However, since the transparency of the BK7 substrate, if the metal film on the BK7 is not thick enough to form opaque sample, the TDTR scan of such sample will cause the wave to the signal, which is a great interference to the PSA result. We apply the Four-Point Probe (FPP) technique to directly measure the resistance (V/I) of the deposited metal film. Then the sheet resistivity of the metal film is

$$\rho_s = C \cdot \frac{V}{I}$$

(10)

where $C$ is the correction factor of FPP, which is dependent on the surface area sizeof the metal film. $C$ of a piece of metal film can be evaluated by measuring the length and the width of the sample surface and referring to the table from [22]. For each metal target, 10 minutes of deposition is made on the $SiO_2$ substrate at 6 mTorr chamber pressure. The size of the metallic film is measured and a TDTR scan with 5X magnification lens is operated. In the following Table 4, we recorded all the targets we have used for the superlattice samples.

The values we have measured are much lower than the thermal conductivity of these metals commonly known. For example, the Al targets we used have only 20% to 40% the thermal conductivity of the numbers from literature. We are not clear what factor causes this low thermal conductivity of the metals. But this might be due to the following

1. The chamber is in frequent use with different targets, and this may bring impurity to any depositions.

2. The chamber base pressure is not low enough (lower than $10^{-7}$ mTorr) when we deposit these metal films.

3. The plasma intensity on the sample is high enough to disrupt crystallization of the films.

2.6. SEM Analysis

We use multiple technologies to get the data needed for analysis of the thermal measurements. First, we bring the superlattice sample on silicon substrates for SEM measurement. The SEM requires that all the component of the target be non-insulators so that the electron beams are able to subject to the sample [23, 24, 25, 26, 27, 28, 29, 30, 31, 32]. For this reason, we always simultaneously made an additional superlattice on silicon substrates. Prior to SEM measurement, we cleave the silicon-substrate-based samples, placing the sample at a 90-degree holder to let the cross section of the sample facing electron lens source. This enables the electronic beam to reach the cross section perpendicularly to output the image of the sample without any tilting angle. We use SEM to determine the sputtered film thickness. Unfortunately, the best resolution for SEM is about 3 nm, which is does limit our ability to distinguishing superlattice layers from each other.

**3. Results and Discussions**

In the following SEM images (Figure 2a and 2b), we determine the total sample thickness (including the superlattice structures and top Al layer) of 440 nmand 677 nm for 5 nm superlattice sample and 10 nm superlattice sample, respectively. Because of the relatively insufficient resolution, we are not able to see the superlattice multilayer configurations. In some samples, we appear to see porous structures of the superlattice layers. The reason of such existence is still not known. The Al transducer layer is coated on the top of the surface. We further confirm the existence of Al top layer by looking at the Energy-disperse X-ray spectroscopy (EDX), which correspondingly scans the targeted surface to analyze the existing elements by securing peaks of differentelement. Figure 3(a-c) show the EDX scan of single elements within the 5nm superlattice sample. A higher concentration of an element will result in more concentrated dots in the EDX image. The sample surface is facing toward top-left direction. Thus, from the left-hand side to the right-hand side of these images, the component should start from Al, which contributes to the top layer way thicker than each superlattice layer. In the middle area there are superlattice bilayers consisting of both Al and Cu. For a larger area to the right, it is silicon substrate. These imageshelp with figuring out the concentration and distribution of the metals and silicon at their corresponding places. We confirm the existence of the Al transducer on the top because it is obvious that, in Figure 3a, there is a bright and concentrated aqua color ribbon which is less than 100 nm wide. Due to the pure Al layer deposited, the concentration of Al in this ribbon region is much higher than its following area on the right side, where Al partially composes of the superlattice layers. The less bright regionon the left side of the ribbon is the sample surface. The reason it can be projectedon the image is that the sample is not placed perfectly at 90 degrees. Thus, because of slight tilting, the surface of the sample is showing in the EDX image. For the second image

(Figure 3b), it only shows a red color region which overlaps with the same region with Al in Figure 3a. It is confirmed that this area is the superlattice region. This image explains that Cu is the other element at this superlattice region except Al. For the last image (Figure 3c), we see a large area of green color which comes from the silicon substrate. The less concentrated area within Figure 3 c is due to the noise of the EDX and there are also noises in the first two images where both metals seemed "diffused" into the silicon substrate.

Because we are unable to testify the existence of the superlattice structureby SEM and EDX. We must use other techniques to demonstrate the superlattice configurations within the sample. Here in Figure 4, we have scanned 10 nm Al/Cu superlattice sample using picosecond acoustics two times, first one with 35 mW pump power and 14 mW probe power, and second one with 77 mW pump power and 9 mW probe power. We take two scans with different output powers of the pump and the probe beams to make sure the signal peaks are not noises. If there are acoustic response of sample structures, they should appear on the graph at the same time delay with whatever laser input power. We use a 10X magnification and it has a focal length of 200 mm. The spot size measured by a laser CCD camera (Spiricon SP620U) is 2.718mm in diameter. Using Eq. (1), we determine a spot size of 18.4 $\mu$m (radius). The spot size, according to the sensitivity plots, has no significant influence on the TDTR measurement. From the picture, the first sharp peak right after time delay isthe thermal acoustic echo of the interface between top Al transducer layer and the upper layer of Cu within superlattice structures. Because the acoustic impedance of Al is lower than Cu, the echo peak appears on the figure in the form of an upwardpeak with a 'tip'. As late as around $t = 247$ ps, another upward peak is witnessed.This peak stands for the Al/sapphire interface echo, where the Al is the lower and the first deposited layer of the superlattice structures on sapphire substrate. Because the impedance of Al is also smaller than that of sapphire, we see an upward "tip" peak again.

We then operate the more accurate PSA analysis (Figure 5). With the Al/Cu peak happening at 16.84 ps and the Al/Sapphire happening 229.73 ps after the Al/Cupeak. The first calculation is the Al thickness. With the known deposition rate of Al is 5 nm/min, the deposition rate, accordingto the thickness and deposition time, is 4.5 nm/min. This situation is acceptable asthe deposition of a target in our sputtering chamber may fluctuate with changes inbase pressure, temperature, target shape and consumption, as well as the reflected RF power. Additionally, there are a few seconds delay of the shutter in response to the switching instruction from the system. In the following depositions after Cu samples, we raised our power of RF from 180W to 195W, making the actual deposition rate closer to 5 nm/min. With the SEM result of a total thickness of the 10 nm Cu sample 677 nm, we have deducted the Al top layer thickness and finally get the thickness of the superlattice layer, 623 nm. This superlattice structure has 28 bilayers and thus, each bilayer is 22.25 nm thick. The Al layer of the Al/Cu bilayer is 9 nm, based onthe calculated deposition rate of Al. Thus, the Cu layer is 13.25 nm thick. The range between the actual layer thickness of Al and Cu (9 nm and 13.5 nm) is not in great difference from the anticipated thicknesses (10 nm + 10 nm).

We carry out the TDTR measurements and data analysis of the superlatticesamples. To run the analysis in our MATLAB program we require known values from the samples of the thermal conductivity, volumetric heat capacity and thickness of eachlayer for all layers except the one which we are fitting (for that layer an initial guessis all that is required). With the initial value and the other input data, the program creates a series of simulation curves which is compares to the experimental data. It changes the initial guess until it minimized the least squares residual between the experimental and simulated curves. With the output value (in our case, the effective thermal conductivity of the superlattice), we take this calculated effective thermal conductivity into Eq. (3) and reversely calculate the $G$, the interfacial thermal conductance.

As for the choice of $G$ as the initial condition for the determination of the effective thermal conductivity, we take the value simulated by the diffuse mismatch model raised in Gundrum's work. The value is referred from Table 2. Also, as a verification, we take the measured Al/Cu interface conductance $4 GW/m^2 \cdot K$ We want to verify that the Al/Cu interfacial conductance is in consistent to the result in that paper. We firstly fit the TDTR scan data for 10 nm Al/Cu superlattice. We analyze the data points for the time delay range from 100 ps to 3700 ps. According to the data, we see a matching between the experimental points and the simulated curve. The simulation curve made by the program is in well alignment with the experimental points The 10nm sample have a total thickness of 677 nm (SEM result), with Al toptransducer of 54 nm and 623 nm superlattice thickness. Each Al is 9 nm and Cu is 13.25 nm. There are 28 bilayers in the superlattice structure. I used a spot size of 5 $\mu$m (10X lens, this is an estimated value after I discussed with Rasel and we need to do beamoffset measurement for the actual spot size) and total laser power of 84.4 mW fittingthe curve above and it gives me a calculated=41.9GW/m$^2$ ·K. The back-calculated $G$ regarding the calculated is 4.42GW/m$^2$K. We take the thermal conductivity of Aland Cu as $\kappa_{Al}$ = 207W/m · K and $\kappa_{Cu}$= 370W/m · K. However, this result may not be trustworthy as the measurement has a great sensitivity to the spot size when it isless than 10 $\mu$m. Also, the thermal conductivity of the metals we took for this scan is from the literature values. Thus, to measure the metal-metal interfacial conductance of the superlattice layers more accurately, we applied the 2X lens and measure the direct value of the metal target by FPP method.

For the analysis of all the samples, we use 2X lens as the laser focusing magnification. This is because that larger beam spot has less influence on the TDTR fitting (less sensitivity) and thus, with 2X lens, we can get more accurate results than 5X and 10X lenses. Figure 7 demonstrates the advantage of 2X magnification lens over other lenses for TDTR measurement. The sensitivity of 2X in response to time delay is much less significant than the other two lenses, giving the influence to the result that can be negligible. This means that 2X is preferred in focusing the laser beam. Despite 2X lens giving weaker signal than 5X and 10X lenses, the aluminum top layer has high enough thermoreflectance coefficient to allow for visible enough signal for all the lens above.

For two different Al/Cu samples, we see different outcomes after we took TDTR fitting to the data. For 10 nm Al/Cu superlattice sample, using $4 GW/m^2 \cdot K$ as the initial fitting interfacial thermal conductance, we can determine an interfacial thermal conductance of $4 GW/m2 \cdot K$. As there are distinguishable superlattice echoes found in the TDTR scan of the 10 nm sample, there are integrated multilayer structures in 10 nm superlattice configurations. Figure 6 illustrates the fitting curve of the 10 Al/Cu sample and the actual data points. To calculate the interface thermal conductance of the superlattice layers, the thickness must be known accurately. For normal single film, we may use PSA to determine the thickness of the film layer at great accuracy. However, since the sound propagation in superlattice layers is more complicated than single film, it is difficult to determine thickness by PSA. First, the exact ratio between the Al and X metal layer in the superlattice configuration is not precisely known. The effective acoustic speed in the superlattice configuration cannot be calculated if the thickness of each aluminum and the other metal is not accurately known. In addition, the acoustic wave goes through several transmissions and reflections across superlattice layer interfaces, which may cause changes to the acoustic speed. With such limitations, we are not able to measure the thickness of the deposition film on the substrate by PSA. Instead, we use the more direct and accurate method, SEM, to measure and determine the thickness of all the samples.

From Figures 8-13, we use SEM to image the cross-section region of the superlattice samples. This allows us to determine the physical thickness of the superlattice samples. With the thickness known, we apply the measured thermal conductivity values along with the thickness of each layer, to fit the TDTR curve with the MATLAB code. The code firstly takes all the parameters along with a value we estimated the effective thermal conductivity of superlattice to be, calculates and plots TDTR curve and makes corrections based on the overlapping between the simulated curve and the experimental data curve and finally gives out a most overlapping curve that comes with the calculated effective thermal conductivity. All the TDTR fittings are shown in Figure 14.

Table 5 illustrates the result of TDTR analysis of the superlattice samples. Among all the Al/X metal interfaces, the Al-Cu interface has the greatest $G$ value of 20.92 GW/$m^2$·K, 5 times larger than the value in Gundrum's work. For the rest of the data, they are measured for the first time and there has been no experimental work done to present values. As for the uncertainty and the accuracy of the measured values, we have applied error bars on all the data of the $\kappa_{eff}$ and calculated $G$ values. The error bars of the $\kappa_{eff}$ is directly calculated by the MATLAB code along with the value of the $\kappa_{eff}$, with an upper limit and a lower limit.

For the error bar of the $G$ value, we use Eq. (5) as the target for the uncertainty calculation and use the general uncertainty function.

With a further comparison, Figure 15 illustrates the distribution of the experimental data and error bars of $G$ in terms of the Z value of the metal forming superlattice with Al. We show the Z-G points as solid squares, which lie on the $G_{Al=X}$ curve

$$\frac{7.705 \times 10^{10} Z_x}{4(7.705 \times 10^{10} + Z_x)} \qquad (11)$$

In this $Z_x$-$G$ curve, the $7.705 \times 10^{10}$ value is directly referred from the Z value of aluminum based on DFT model at 300K. Among the six samples we have mea- sured $G$ with, the 5nm/5nm Al/Cu sample gave us the data point closest to the $Z_x$-$G$ curve. However, we found that the thermal conductance of 10 nm/10 nm Al/Cu interface is much larger than the value of 5nm/5nm. Despite such a distinct difference, the error bars of these Al/Cu interfacial thermal conductance data have overlapping. Thus, we assume that the interfacial conductance of Al/Cu layer may likely be setting between 8.51 GW/m² ·K (the lower limit of the 10 nm/10 nm Al/Cu error bar) and 8.81GW/m² · K (the upper limit of the 5 nm/5 nm Al/Cu error bar). For Al/Ag and Al/Ni interfaces, their interfacial thermal conductance values are much less than what DFT model predicted. We see a quite uniform conductance values between 5 nm/5nm Al/Ni and 10 nm/10 nm Al/Ni interfaces. There is also an overlapping between the measured thermal conductance of the two types of Al/Ni interfaces considering the range of the error bars.

## 4. Conclusions

In this study, the interface thermal conductance of Al/Cu, Al/Ni, Al/Fe and Al/Ag in different superlattice layer configurations are measured. These conductance values were much larger than previous work in terms of Al/Cu interface. The values measured do not agree particularly well with the density functional theory model in the G-$Z_x$ graph.

**Acknowledgement**

D.L. greatly thanks Dr. Yuying Zhang for the help in collecting SEM data.

**Figure Captions**

Table 1. List of Thermal Property of Tested Metals (Green color filled metals have the highest sensitivities, orange color have medium and yellow color has low sensitivity. Al is the thermal transducer coated over all the metals above to determine thermal conductance)

| Metal Name | $D$ (m$^2$/s) | $L_p$ (nm) | $\kappa$ (W/m·K) | $C_p$ (J/kg·K) | $C_v$ (J/m$^3$·K) |
|---|---|---|---|---|---|
| Cu | 1.11e-4 | 1675 | 401 | 376 | 3.37 |
| Ag | 1.66e-4 | 2048 | 427 | 238 | 2.50 |
| Co | 2.79e-5 | 840 | 100 | 418 | 3.70 |
| Ni | 2.40e-5 | 779 | 91 | 502 | 4.47 |
| Cr | 2.87e-5 | 852 | 94 | 460 | 3.29 |
| Fe | 2.30e-5 | 762 | 80 | 460 | 3.62 |
| Ti | 9.20e-6 | 482 | 22 | 544 | 2.45 |

Table 2. Free electronic heat capacity properties of metals and the calculated inter-facial conductance with Al by the free electron model.

| | $n$ | $E_f$(eV) | $V_f$(m/s) | $T_f$(K) | $\gamma T$(J/m$^3$·K) | $Z$(J/m$^2$·K·s) | $G$(W/m$^2$·K) |
|---|---|---|---|---|---|---|---|
| Al | 1.81E+29 | 11.65 | 2.02E+06 | 1.36E+05 | 2.73E+04 | 1.83E+10 | |
| Cu | 8.48E+28 | 7.03 | 1.57E+06 | 8.16E+04 | 2.12E+04 | 3.33E+10 | 5.20E+09 |
| Ag | 5.86E+28 | 5.49 | 1.39E+06 | 6.38E+04 | 1.87E+04 | 2.61E+10 | 4.43E+09 |
| Fe | 1.7E+29 | 11.10 | 1.98E+06 | 1.30E+05 | 2.67E+04 | 5.19E+10 | 6.76E+09 |
| Sn | 1.48E+29 | 10.03 | 1.88E+06 | 1.18E+05 | 2.55E+04 | 4.86E+10 | 6.45E+09 |
| Ni | 1.83E+29 | 13.97 | 2.22E+06 | 1.36E+05 | 2.74E+04 | 6.70E+10 | 6.93E+09 |

Table 3. Effective thermal properties and layer physical configuration of Al/X super-lattice samples.

| Al/X | $\kappa_{eff}$(W/m·K) | $D_{eff}$(m²/s) | $L_{P\,s}$(nm) | # of bilayers | $t\,s$(nm) |
| --- | --- | --- | --- | --- | --- |
| Al/Cu (5 nm) | 18.71 | 4.89E-06 | 352 | 42 | 420 |
| Al/Cu (10 nm) | 35.14 | 9.19E-06 | 482 | 28 | 560 |
| Al/Ag (5 nm) | 18.74 | 4.90E-06 | 352 | 46 | 460 |
| Al/Ag (10 nm) | 35.27 | 9.22E-06 | 483 | 32 | 640 |
| Al/Sn (5 nm) | 16.79 | 4.39E-06 | 333 | 60 | 600 |
| Al/Sn (10 nm) | 28.92 | 7.56E-06 | 437 | 36 | 720 |
| Al/Fe (5 nm) | 17.14 | 4.48E-06 | 337 | 58 | 580 |
| Al/Fe (10 nm) | 29.98 | 7.84E-06 | 445 | 35 | 700 |
| Al/Ni (5 nm) | 17.34 | 4.53E-06 | 339 | 53 | 530 |
| Al/Ni (10 nm) | 30.61 | 8.00E-06 | 450 | 32 | 640 |

Table 4. Metal target thermal properties on SiO₂ substrate and their depositionconditions

| Target | Size (mm) | $C \times R$ | $\kappa$(W/m·K) | deposition rate (nm/min) | Pwr & conditon |
| --- | --- | --- | --- | --- | --- |
| 99.99%Al | 7 × 7 | 3.9 × 0.58 | 54 | 6.1 | 120W@DC |
| 99.99%Cu | 6 × 4.5 | 3.34 × 0.126 | 140 | 12.4 | 180W@RF |
| Fe | 8.5 × 4.5 | 3.38 × 5.91 | 10 | 5.5 | 180W@RF |
| 99%Al+1%Si | 9 × 4.5 | 3.4 × 0.376 | 93.86 | 6.1 | 120W@DC |
| 99.99%Ag | 7.5 × 6 | 3.72 × 0.226 | 277 | 31.4 | 180W@RF |
| 99.99%Ni | 7.5 × 6 | 3.72×1.073 | 24.3 | 7.55 | 180W@RF |

Table 5. Physical conditions of the superlattice metal samples along with the measured superlattice effective thermal conductivity and the interfacial conductance.

| Sample | Total Thickness (nm) | $\kappa_{eff}$(W/m·K) | $G$(GW/m²·K) | Layer Configuration |
|---|---|---|---|---|
| Al/Cu | 596 | $56.4(^{+6.2}_{-2.2})$ | $20.92(\pm 12.41)$ | 10 nm + 10 nm |
| Al/Cu | 447 | $23.47(^{+3.02}_{-3.26})$ | $7.27(\pm 1.54)$ | 5 nm + 5 nm |
| Al/Ag | 705 | $24.96(^{+4.06}_{-1.52})$ | $4.20(\pm 0.57)$ | 5 nm + 10 nm |
| Al/Ni | 649 | $11.14(^{+1.65}_{-1.51})$ | $1.71(\pm 0.3)$ | 10 nm + 10 nm |
| Al/Ni | 516 | $8.75(^{+1.75}_{-1.19})$ | $2.64(\pm 0.72)$ | 5 nm + 5 nm |
| Al/Fe | 691 | $15.12(^{+1.57}_{-1.41})$ | $16.10(^{+30}_{-16.1})$ | 10 nm + 10 nm |

Table 6. the comparison of thermal conductance between the diffuse mismatch model and the density functional theory model and the actual measured values

| Interface(Al-X) | $v_{XF}$ ($10^6$m/s) | $\gamma_{X\,expt}$ (J/m²·K²) | $G_{dmm}$ (GW/m²·K) | $G_{DFT}$ (GW/m²·K) | $G_{expt}$ (GW/m²·K) | $C_{eX\,expt}$ (J/m²·K) |
|---|---|---|---|---|---|---|
| Al-Cu (10 nm) | 1.57 | 105.10 | 6.09 | 7.819 | $20.92(\pm 12.41)$ | 3.15E+04 |
| Al-Cu (5 nm) | 1.57 | 105.10 | 6.09 | 7.819 | $7.27(\pm 1.54)$ | 3.15E+04 |
| Al-Ag(5-10 nm) | 1.39 | 65.27 | 4.34 | 6.860 | $4.20(\pm 0.57)$ | 1.96E+04 |
| Al-Fe(10 nm) | 1.98 | 1169.97 | 11.20 | 10.76 | $16.10(^{+30}_{-16.1})$ | 3.51E+05 |
| Al-Ni(10 nm) | 2.22 | 1229.53 | 11.30 | 11.25 | $1.71(\pm 0.3)$ | 3.69E+05 |
| Al-Ni(5 nm) | 2.22 | 1229.53 | 11.30 | 11.25 | $2.64(\pm 0.72)$ | 3.69E+05 |
| Al | 2.03 | 78.86 | N/A | N/A | N/A | 2.37E+04 |

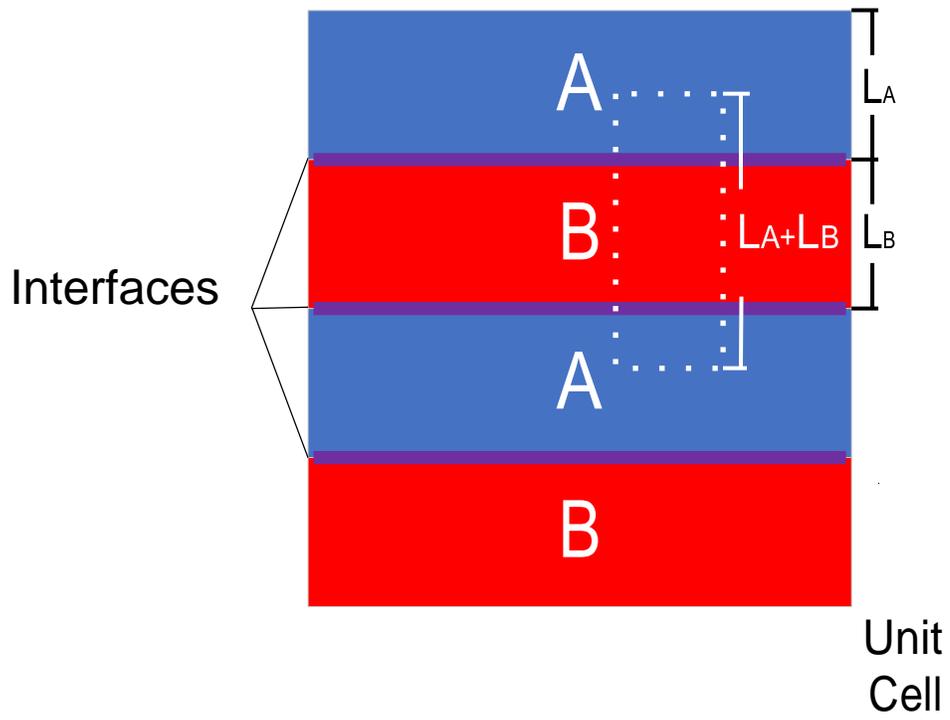

Figure 1. The unit cell of an example superlattice layers

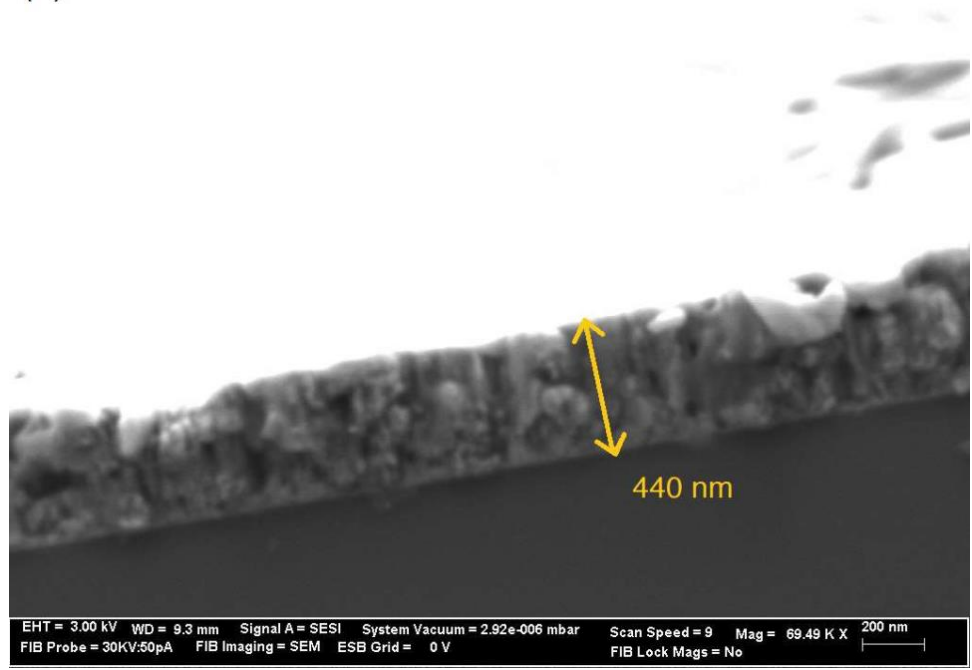

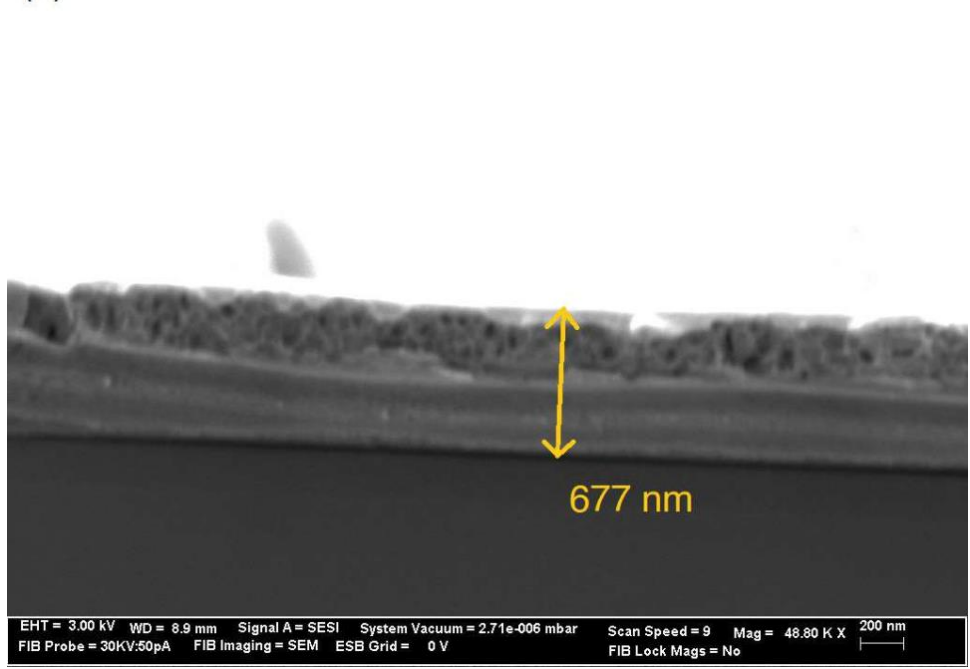

Figure 2. SEM images of Al/Cu superlattice sample cross sections. (a). 42X 5 nmAl/Cu superlattice bilayers; (b). 28X 10 nm Al/Cu superlattice bilayers.

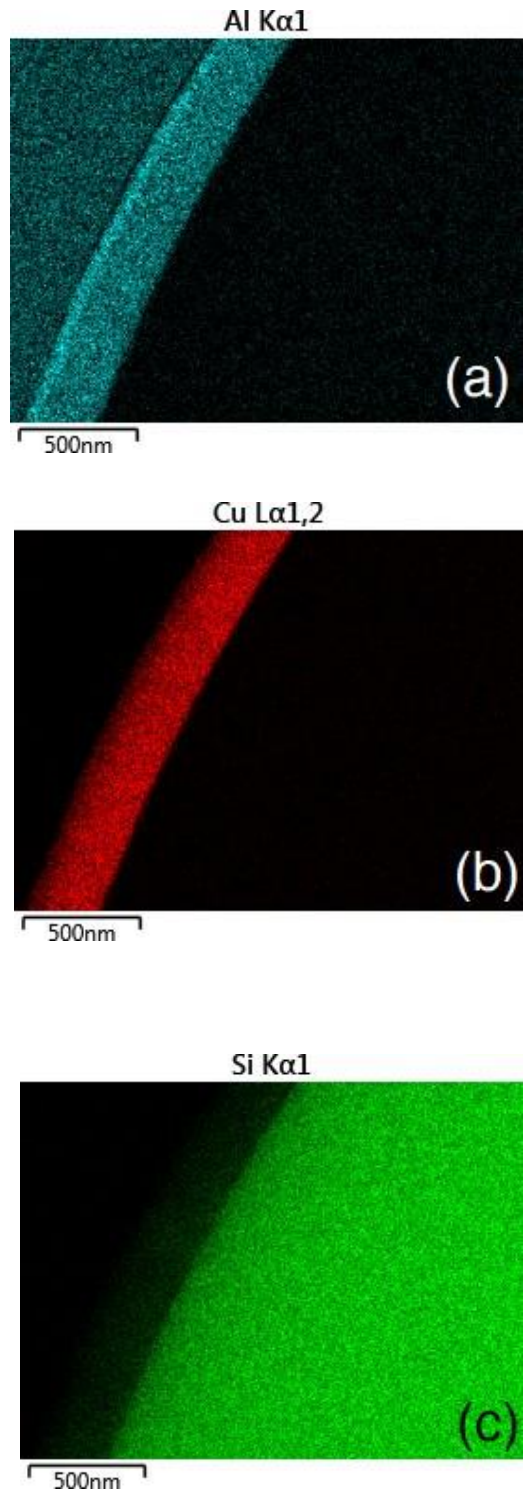

Figure 3. EDX images of Al/Cu superlattice sample cross sections (silicon substrate).(a). Al element concentration; (b). Cu element concentration; (c). Si element concentration

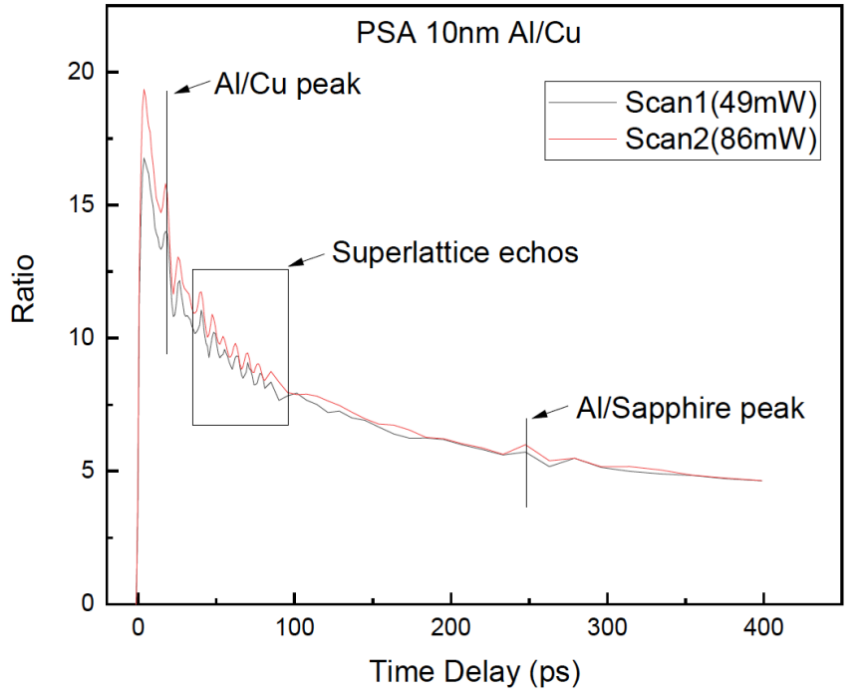

Figure 4. TDTR scan of Al/Cu superlattice sample on sapphire substrate

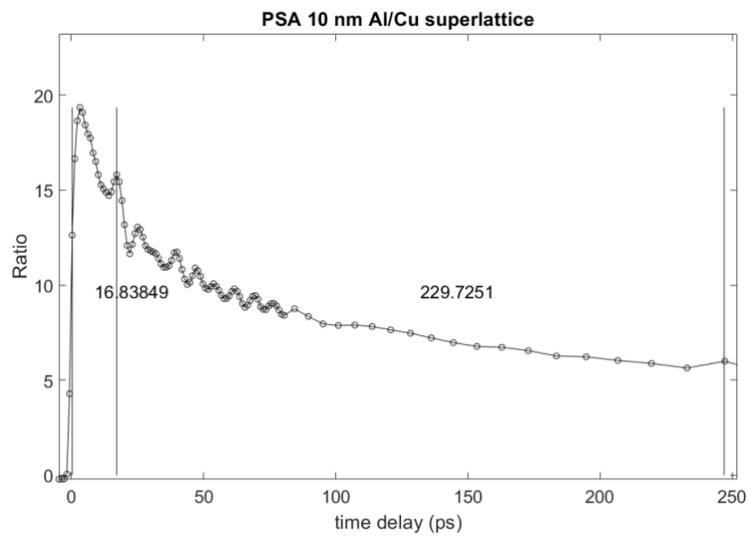

Figure 5. PSA analysis of Al/Cu superlattice sample on sapphire substrate

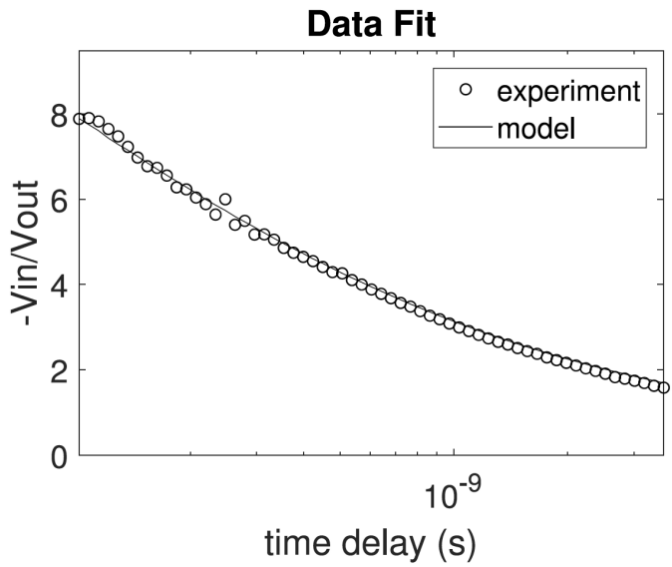

Figure 6. TDTR fitting of the 10 nm Al/Cu superlattice (623 nm) experimental dataat 77mW pump power and 7.4mW probe power.

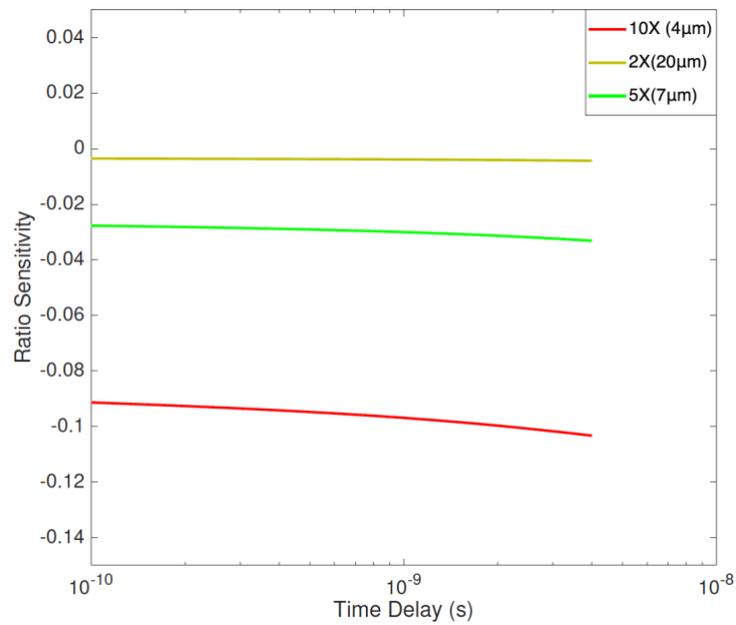

Figure 7. Sensitivity plot of the TDTR experiment using 2X, 5X and 10X magnification lens

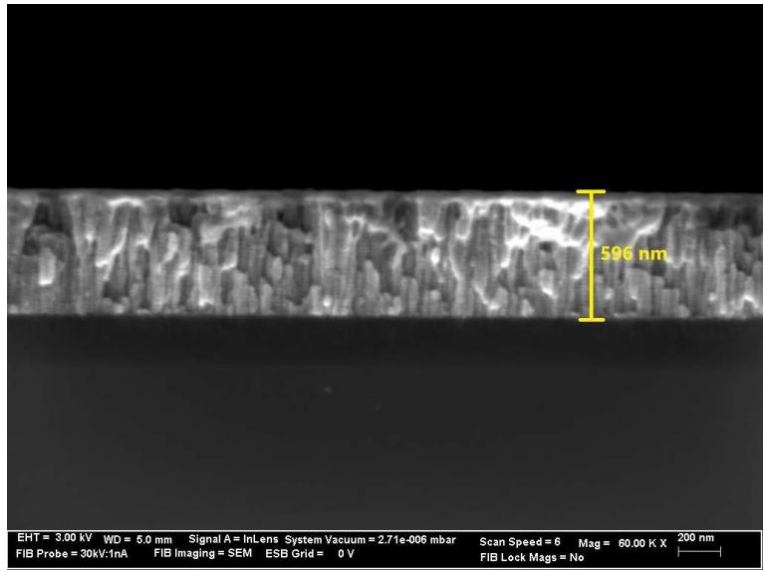

Figure 8. SEM image of the cross section of 10nm Al/10nm Cu superlattice sample

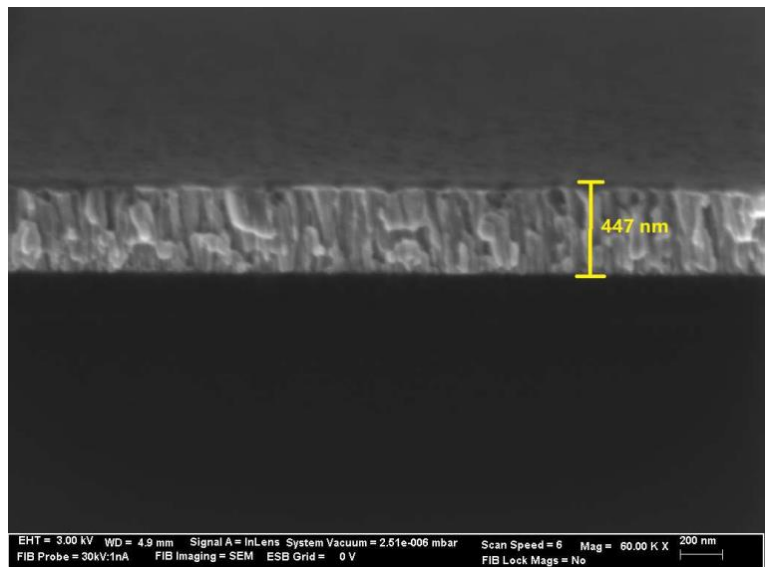

Figure 9. SEM image of the cross section of 5nm Al/5nm Cu superlattice sample

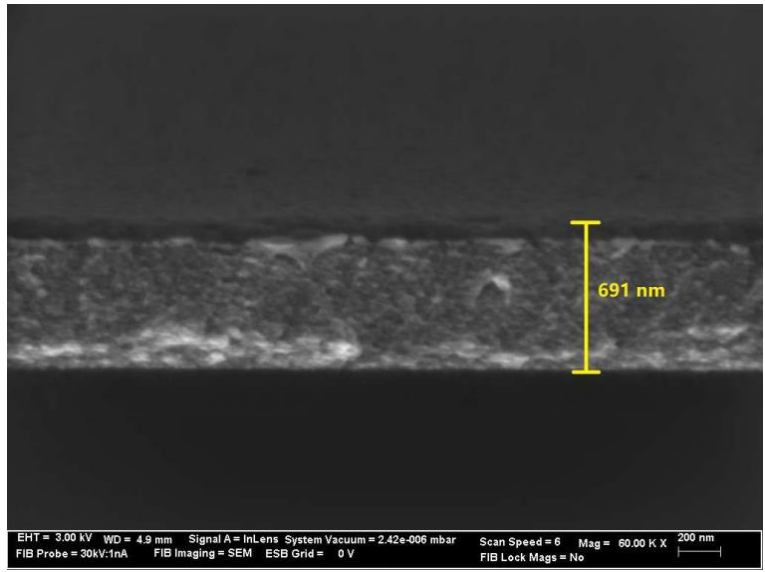

Figure 10. SEM image of the cross section of 10nm Al/10nm Fe superlattice sample

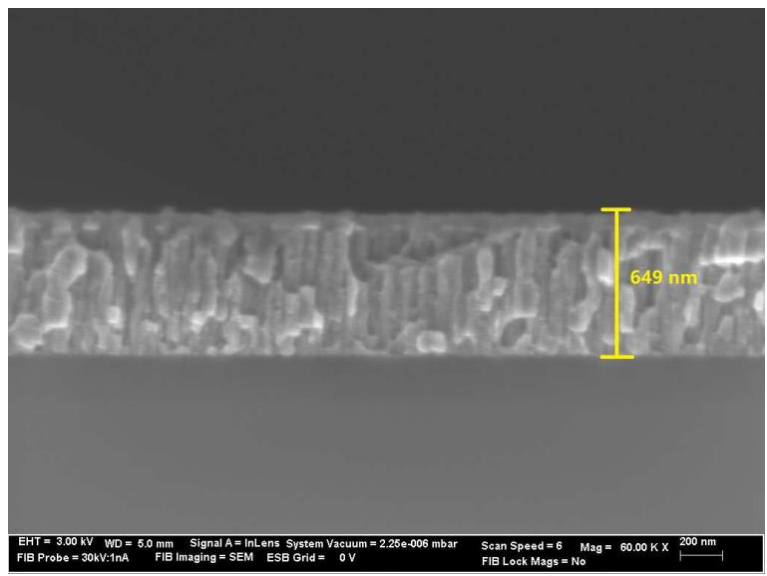

Figure 11. SEM image of the cross section of 10nm Al/10nm Ni superlattice sample.

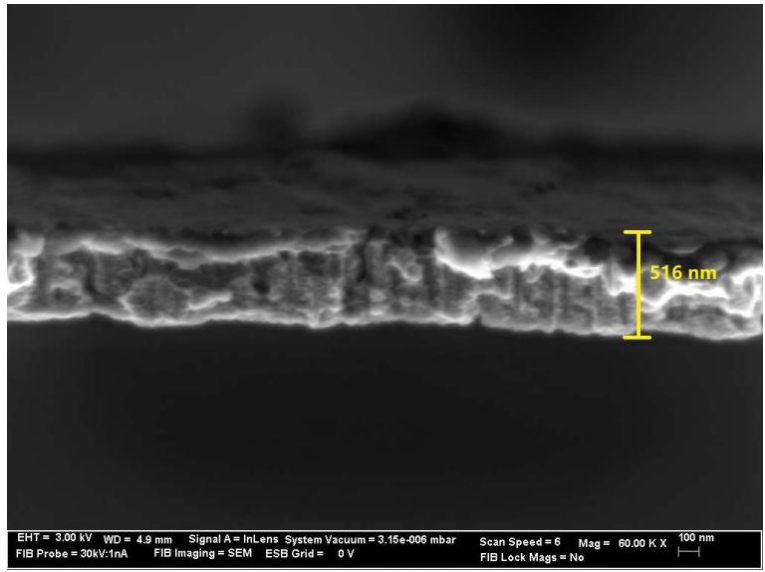

Figure 12. SEM image of the cross section of 5nm Al/5nm Ni superlattice sample

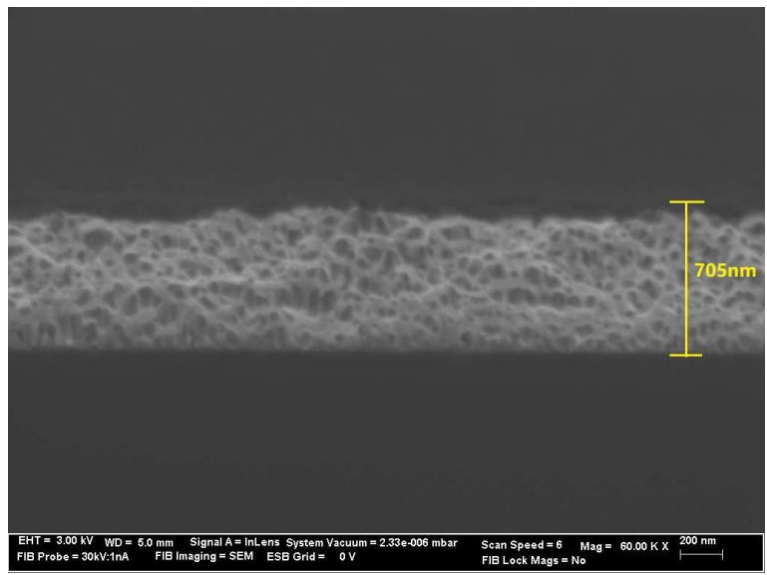

Figure 13. SEM image of the cross section of 5nm Al/10nm Ag superlattice sample

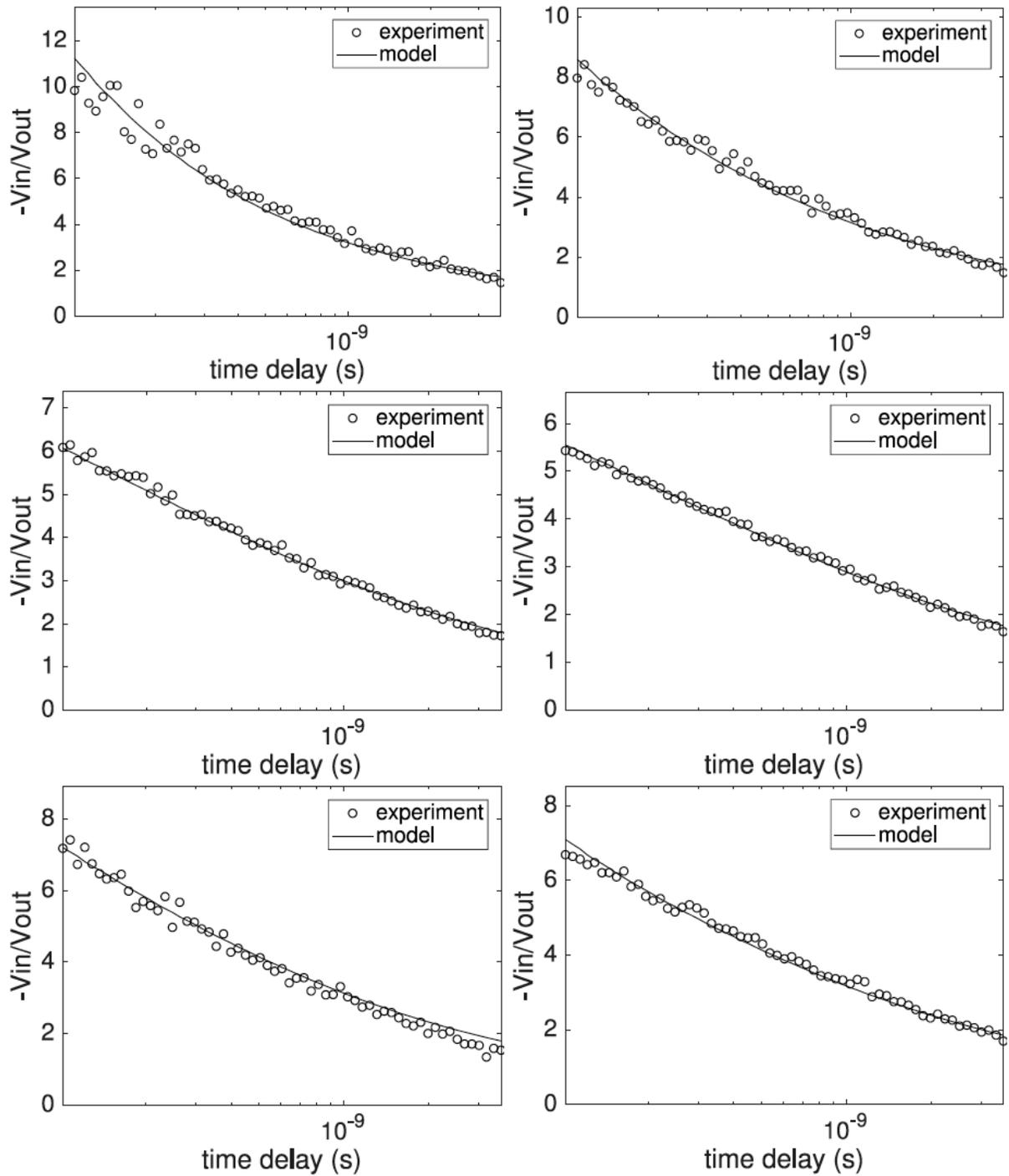

Figure 14. TDTR fitting of the Al/X superlattice experiments. top-left: (10nm Al/10nm Cu)$_{28}$, top-right: (5nm Al/5nm Cu)$_{42}$, middle-left: (10nm Al/10nm Ni)$_{32}$, middle-right: (5nm Al/5nm Ni)$_{53}$, bottom-left: (5nm Al/10nm Ag)$_{46}$, bottom-right: (10nm Al/10nm Fe)$_{35}$

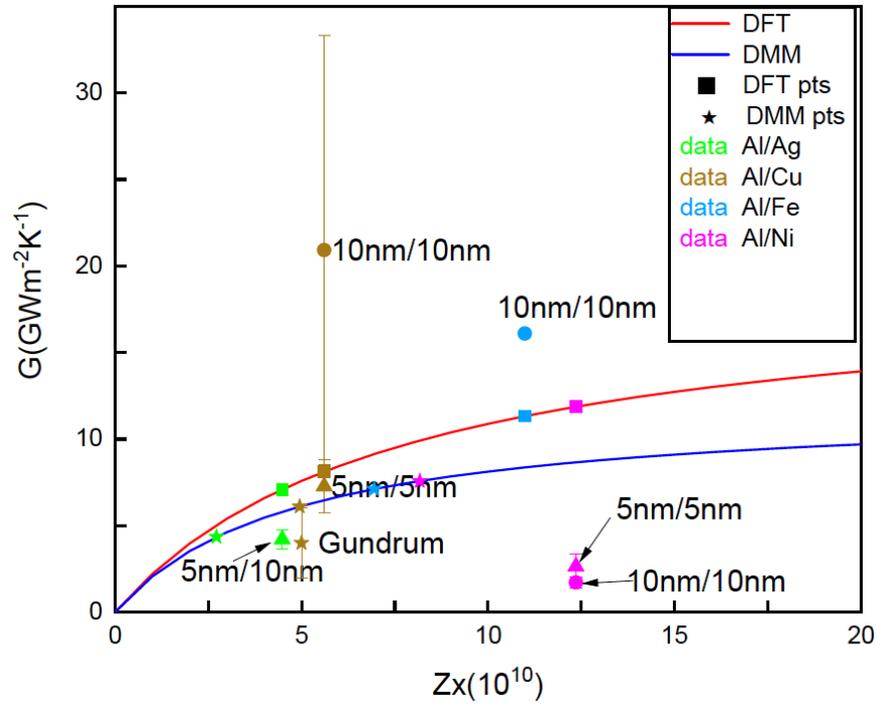

Figure 15. The interfacial conductance data in the form of $Zx$-$G$ chart (the measured Al/Cu interfacial conductance value in Gundrum's work is recorded as a brown color star in this figure)